\documentclass[11pt,a4paper]{article} 
\usepackage{amsmath}
\usepackage{graphicx}

\usepackage{amssymb}
\usepackage{authblk}

\title{ $217\ \mathrm{km}$ long distance photon-counting optical time-domain reflectometry  based on ultra-low noise up-conversion single photon detector  }

\author[1,$\dagger$]{Guo-Liang Shentu}
\author[1,2,$\dagger$]{Qi-Chao Sun}
\author[1,$\dagger$]{Xiao Jiang}
\author[3]{Xiao-Dong Wang}
\author[4]{Jason S. Pelc}
\author[4]{M. M. Fejer}
\author[1,*]{Qiang Zhang}
\author[1]{Jian-Wei Pan}

\affil[1]{Shanghai Branch, National Laboratory for Physical Sciences at Microscale and Department of Modern Physics,University of Science and Technology of China, Shanghai, 201315, China}
\affil[2]{Department of Physics, Shanghai Jiao Tong University, Shanghai, 200240, China}
\affil[3]{College of Physics and Electronic Engineering of Northwest Normal University, Lanzhou, 730070, China}
\affil[4]{Edward L. Ginzton Laboratory, Stanford University, Stanford, California, 94305, USA}
\affil[$\dagger$]{These authors contributed equally to this work}
\affil[*]{Corresponding author: qiangzh@ustc.edu.cn}

\date{\today}
\begin{document}
\maketitle
We demonstrate a photon-counting  optical time-domain reflectometry with $42.19\ \mathrm{dB}$ dynamic range using an ultra-low noise up-conversion single photon detector. By employing the long wave pump technique and a volume Bragg grating, we reduce the noise of our up-conversion single photon detector, and achieve a noise equivalent power of $-139.7 \ \mathrm{dBm/\sqrt{Hz}}$. We perform the OTDR experiments using a fiber of length $216.95\ \mathrm{km}$, and show that our system can identify defects along the entire fiber length with a distance resolution better than $10\ \mathrm{cm}$ in a measurement time of $13 \ \mathrm{minutes}$.

\section{Introduction}

Optical  time-domain reflectometry (OTDR) is a commonly used measurement technique for fiber network diagnosis. By detecting the Rayleigh backscattered light of a pulse launched into fiber under test (FUT), one can get information about the attenuation properties, loss and refractive index changes in the FUT \cite{conventionalotdr£ºBarnoski,conventionalotdr£ºPersonick}. Conventional OTDRs using linear photodetectors are widely used, but their performance is limited by the high noise equivalent power (NEP) of the p-i-n or avalanche photodiodes used in these systems. Photon-counting OTDRs ($\nu$-OTDR), which employ  single photon detectors instead, have been the subject of increased attention, becauase they offer better sensitivity, superior spatial resolution, an inherent flexibility in the trade-off between acquisition time and spatial precision, and the absence of the so-called classical dead zones.

Several $\nu$-OTDR systems have been demonstrated with InGaAs/InP avalanche photodiode (APD) operated in Geiger-mode \cite{InGaAsOTDR:44db,InGaAsOTDR:Wegmuller}. But in these demonstrations, the InGaAS/InP APDs used suffer from noise issues caused by large dark current and after pulsing \cite{InGaAsOTDR:limit}. Time gated operation of the detectors is used in these $\nu$-OTDR measurements to reduce the noise. However, a consequence of time gated operation is that it only allows part of the fiber to be measured at a time, so the measurement time is longer than that using  free running detectors by almost 3-order of magnitude \cite{upconversion:Hirokie}. Recently, $\nu$-OTDRs based on free running superconducting single photon detectors (SSPD) have been reported \cite{SSPD:Hu,SSPD:Tang}. Thanks to the low NEP of SSPD, which is about $-140.97 \ \mathrm{dBm/\sqrt{Hz}}$, a dynamic range of $37.4 \ \mathrm{dB}$ is achieved in a total measurement time of about $10\ \mathrm{minutes}$  \cite{SSPD:Tang}. However, the superconducting nanowire is operated in a bulky liquid helium cryostat to reduce the thermal noise. Up-conversion single photon detectors that consist of a frequency upconversion stage in a nonlinear crystal followed by detection using a silicon APD (SAPD), provide an elegant room-temperature free-running single-photon detection technology, and have been successfully applied in $\nu$-OTDR systems \cite{upconversion:Hirokie,upconversion:Gisin}. Recent results include a two-point resolution of $1\ \mathrm{cm}$ \cite{upconversion:Gisin} and a measurement time more than $600$ times shorter  \cite{upconversion:Hirokie}. But these systems are not appropriate for long-distance fiber measurements, for the  NEP of these up-conversion single photon detectors is  about 2-order of magnitude larger than that of the SSPD. The high NEP means that much longer measurement times are required to obtain the same signal-to-noise (SNR) ratio at the end of the $\nu$-OTDR trace.

In a recent paper, we demonstrated that the up conversion single photon detector by using long wavelength pump technology, and a volume Bragg grating (VBG) as a narrow band filter to suppress the noise\cite{upconversion:Shentu}. The up-conversion single photon detector we used in the experiment has a NEP of about $-139.7 \ \mathrm{dBm/\sqrt{Hz}}$. Here, we employ the ultra-low noise up-conversion single photon detector and a high peak power pulsed laser, and present a $\nu$-OTDR over fiber of  $216.95 \ \mathrm{km}$ length. With  measurement time of  $13\ \mathrm{minutes}$, we achieve a distance resolution of $10 \ \mathrm{cm}$ and dynamic range of $42.19\ \mathrm{dB}$.

\section{Experimental Setup}

\begin{figure}[htbp]
\centering\includegraphics[width=12cm]{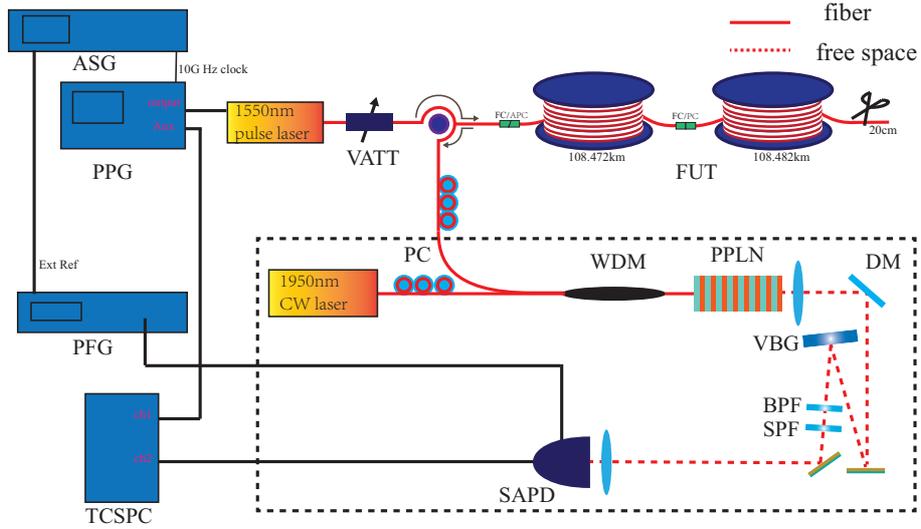}
\caption{(color online) Schematic of the experimental setup. ASG: analog signal generator, PPG: pulse pattern generator, PFG: pulse function arbitrary noise generator, TCSPC: time correlated single-photon counting system, VATT: variable optical attenuator, Circ: optical circulator, FUT: fiber under test, DM: dichroic mirror, PC: polarization controller, SPF: $945\ \mathrm{nm}$ short pass filter, BPF: $857\ \mathrm{nm}$ band pass filter, VBG: volume Bragg grating.}\label{figsetup}
\end{figure}

The experimental setup is shown in Fig.\ref{figsetup}. Laser pulses with central wavelength of $ 1549.87 \ \mathrm{nm}$  are launched into the FUT through an optical circulator. The peak power can be adjusted  using a variable optical attenuator (VATT). The FUT consists of two fiber spools of length $108.47 \ \mathrm{km}$ and $108.48 \ \mathrm{km}$ sequentially. The back scattered light is coupled into the third port of the circulator, and then detected by the ultra-low noise up conversion single photon detector. The output of SAPD is fed into a time correlated single-photon counting system (TCSPC), which is operated in time-tagged time-resolved (TTTR) mode.

An analog signal generator acts as a clock of the whole $\nu$-OTDR system by feeding a  $10\ \mathrm{GHz}$ signal into "clock in" plug of a pulse pattern generator (PPG), and a $10\ \mathrm{MHz}$ signal as external reference of a pulse function arbitrary noise generator (PFG). The PPG's output is used to control the pulse laser, while its auxiliary output connected with TCSPC module to provide a $20\ \mathrm{MHz}$ synchronized clock. The output of PFG is used to switch the SAPD off temporarily  during a repetition period of laser pulse, when we need to measure the FUT by sections.

The up-conversion single photon detector we used for this experiment, shown in the dash box of Fig.\ref{figsetup}, is fully described in \cite{upconversion:Shentu}. The signal light and $1952.39 \ \mathrm{nm}$ pump laser are combined by a $1950\ \mathrm{nm} /1550\ \mathrm{nm}$ WDM and coupled into the z-cut PPLN waveguide through the fiber pigtail. A polarization controller is used to adjust the pump laser to the TM mode, for the PPLN waveguide only supports Type-0 (ee $\to$ e) phase matching. A Peltier  temperature controller is used to keep the waveguide's temperature at $60.8 \ ^{\circ}\mathrm{C}$ to maintain the phase-matching of the sum frequency generation (SFG) process. The generated SFG  photons are collected  by an AR-coated  objective lens, and   separated  from  the  pump  by  a  dichroic  mirror (DM). A VBG, a  $945\ \mathrm{nm}$ short pass filter (SPF) and a  $857\  \mathrm{nm}$ band pass filter (BPF) are used to suppress the noise. Finally, the SFG photons are collected and detected by a SAPD. The dark count rate of the SAPD we used is about $60\ \mathrm{Hz}$. Thanks to the long wavelength pump and the narrow band VBG filter, we can suppress the dark count rate of the up-conversion single photon detector to  $80\ \mathrm{Hz}$ while the detection efficiency is $15\%$, which corresponds to an NEP of about $-139.7 \ \mathrm{dBm/\sqrt{Hz}}$. This condition is set as the operation point in our experiment.

\section{Long distance $\nu$-OTDR Application}

\begin{figure}[htbp]
\centering\includegraphics[width=12cm]{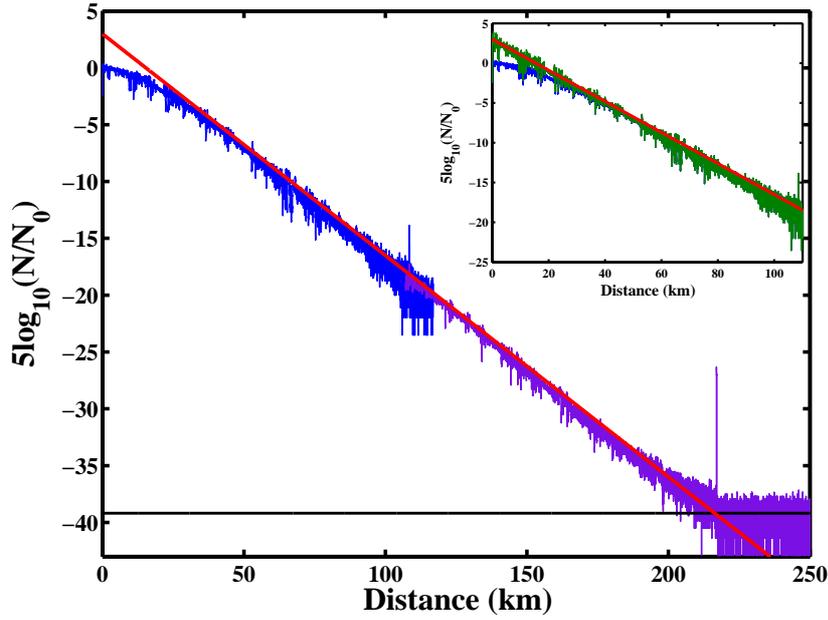}
\caption{(color online) Measurement of optical  fiber of $216.95\ \mathrm{ km}$ length  performed by our $\nu$-OTDR system. The pulse width is  $1\ \mu \mathrm{ s}$. $N$ is the counts of back scattered photons, $N_{0}$ is the count at the the initial point of the trace. The blue trace and violet trace are obtained in the first step and the second step of measurement, respectively. The position of the two peaks, $108.47203 \  \mathrm{km}$ and $108.48219 \ \mathrm{km}$, coincides with the length of the two fiber spools. The black horizontal line shows the RMS noise level of the trace,which is about $-39.19 \ \mathrm{dB}$. The intersection and slope of  the extrapolated trace (red line) are $3\ \mathrm{ dB}$ and $0.195 \ \mathrm{dB/km}$, respectively. The inset shows the comparison between the corrected trace (green) and the trace measured directly (blue).}\label{OTDRtrace}
\end{figure}

In long distance $\nu$-OTDR applications, if the pulse extinction ratio is poor and there is still light in the pulse interval, the back scattered photons of the light will cause non-negligible noise. Therefore the laser pulse extinction ratio is also crucial for a good signal noise ratio. To take advantage of the our low NEP of up-conversion detector, we expect the noise to be below the noise level of the detector. This requires a extremely high pulse extinction ratio of more than $100  \ \mathrm{dB}$.  This is achieved by  providing a small reversed bias voltage to the laser diode; we have confirmed that the emission between laser pulses is below our detection limit . The maximum peak power of our laser pulse is about $23\ \mathrm{dBm}$. The repetition frequency is chosen according to the length of FUT. For a  $216.95 \ \mathrm{km}$-long fiber, the round trip time of laser pulses in it is $2.14 \ \mathrm{ms}$. So the repetition frequency of the laser pulse must be lower than $452 \ \mathrm{Hz}$ and in our experiment, we set the laser pulse repetition frequency at $400 \ \mathrm{Hz}$. The pulse width of the laser is set at $1\ \mu \mathrm{s}$. The measurement is divided  into two steps. The repetition frequency and pulse width are unchanged in the two steps. In the first step, we  perform a  $3\-  \mathrm{minute}$ $\nu$-OTDR measurement and obtain the $\nu$-OTDR trace of the initial $0-120  \ \mathrm{km}$ of FUT. The peak power of the laser is attenuated so as to get a count rate of about $7\times 10^{5}\ \mathrm{Hz}$. And then, we use the maximum peak power, $23\ \mathrm{dBm}$, and perform a $10\  \mathrm{minutes}$ measurement of the remaining fiber. Because the peak power is high, there will be a great amount of photons reflected by the input surface and backscattered by the initial several kilometers of FUT. To protect the SAPD, we switch the SAPD off for  $1\  \mathrm{ms}$ after a pulse is launched into the FUT. Thus, we only get the $\nu$-OTDR trace from $100\ \mathrm{km}$ to the end of FUT in the second step. The two sections are jointed into one according to their time delays, as shown in Fig.\ref{OTDRtrace}.

The up-conversion single photon detector in the experiment is polarization dependent. The strong fluctuation of the $\nu$-OTDR trace in Fig.\ref{OTDRtrace} corresponds to the polarization state revolution when the light propagates through the fiber, and can be used to study the polarization properties of fiber. The polarization induced fluctuation can be eliminated by using a polarization scrambler \cite{upconversion:Gisin} or a polarization independent up conversion single photon detector \cite{polarizationindependent}. The cross talk and Fresnel reflection of the optical circulator will induce a very high peak in the $\nu$-OTDR trace, which is not useful for diagnosing the FUT. Furthermore, the high peak will induce a following dip in the $\nu$-OTDR trace due to the dead time of SAPD and TCSPC. In order to avoid this, we adjust the high peak's polarization so that it has a very low probability to be recorded by our polarization dependent OTDR.

In Fig.\ref{OTDRtrace}, $N$ is the counts of back scattered photons, $N_{0}$ is the count at the the initial point of the trace. Thus $10\log_{10}(N/N_{0})$ represents the total loss in the round trip of the fiber. As is common in OTDR experiments, we plot $5log_{10}(N/N_{0})$, which represent  single-pass loss through the fiber. Considering the $60\ \mathrm{ns}$ dead time of the SPAD, the actual counting rate per time bin should be corrected as, $C_{actual}(t)=\frac{C_{measured}(t)}{1-\sum{C_{mesured}(t^{'})}}$, where $C_{measured}(t)$ is the measured counting rate per time bin, and the summation means the total counting rate during  $80\ \mathrm{ns}$ before the time bin at time $t$. It is obvious that the difference between measured and true actual counting rate is small when the counting rate is very low. In the first step of our experiment, the count rate is more than $7\times 10^{5}\ \mathrm{Hz}$ for the beginning of the $\nu$-OTDR trace. As shown in inset of Fig.\ref{OTDRtrace}, we correct the measurement trace (with a color of blue) with the above formula and the achieved trace (with a color of green) coincidences with the extrapolated trace (with a color of red) obtained by a linear fit of measured trace. The slope of the extrapolated trace indicates the attenuation of fiber of $0.195\ \mathrm{dB/km}$. The intersection of the extrapolated trace is the actual value of the trace at the initial point, which is about $3\ \mathrm{dB}$.  The trace of experiment can be distinguished from the noise obviously at the end of the fiber. The root mean square (RMS) noise level is calculated from the data at the tails of the trace. The dynamic range is about $42.19\ \mathrm{dB}$, which is determined by the difference between the intersection of the extrapolated trace and the RMS noise level.
\begin{figure}[htbp]
\centering\includegraphics[width=12cm]{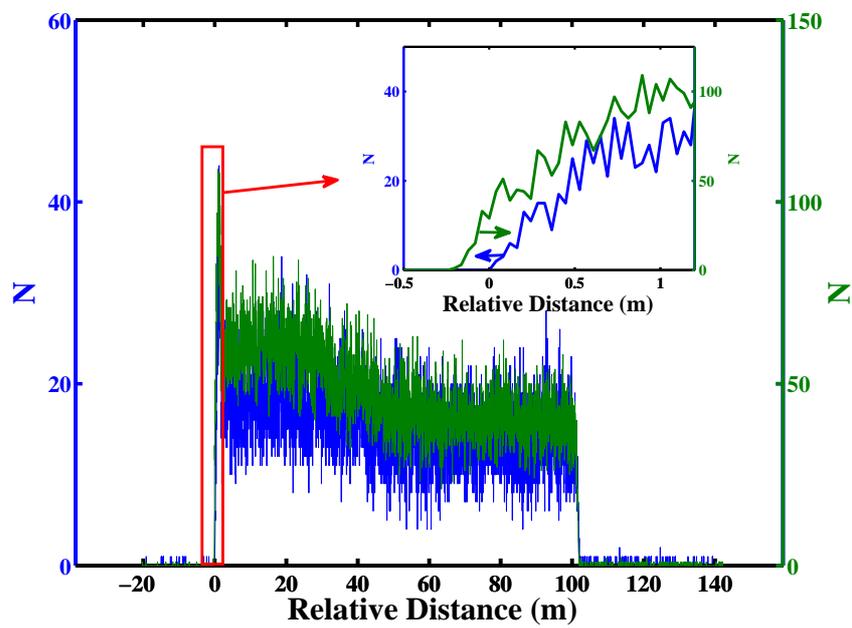}
\caption{(color online) Counts of last reflection peaks of $\nu$-OTDR trace of the $218.95\ \mathrm{km}$ fiber (blue line) and after the $20\ \mathrm{cm}$ fiber is cut off at the end (green line), which are represented by the right y-axis and left y-axis, respectively. The amplitude of the two peaks are different because the cutting surfaces of fiber end are not identical.  The inset shows the enlarged view of the leading edge of the two peaks.  }\label{resolution}
\end{figure}

One important parameter of OTDR is the distance resolution. It is the ability of the OTDR to locate a defect along the FUT. The timing jitter of the detector determines the distance resolution. From our detector timing jitter of $500\ \mathrm{ps}$ we compute a distance resolution of approximately $5\ \mathrm{cm}$. In order to demonstrate the spatial resolution experimentally, we cut $20\ \mathrm{cm}$ fiber off at the end of the second fiber spool, and perform the experiment again as described above. The last reflection peaks of the two $\nu$-OTDR traces are shown in Fig.\ref{resolution}. As shown in the figure, the laser pulse is not broadened after transmitting through $216.95\ \mathrm{km}$ fiber. The leading edges of the two peaks, as shown in inset of Fig.\ref{resolution}, are separated with a 20 cm distance which  coincides with the length of the cutting off fiber. According to the figure, the experimental distance resolution is about $10\ \mathrm{cm}$, which is larger than the expected resolution of $5\ \mathrm{cm}$. The difference is caused by the fluctuation of the counts, which can be improved by extending the measurement time. Note that the distance resolution is different than the two-point resolution, which is minimum distance between the defects can be discriminated. The two-point resolution we can achieve is about $100\ \mathrm{m}$ corresponding to the $1 \ \mu \mathrm{s}$ pulse width we used in our experiment. Using shorter pulses will improve two-point resolution. But meanwhile, shorter pulses with a constant peak power means less photons in the pulse, which will decrease the measuring range and resolution.

\section{Conclusion}
In conclusion, we have presented the implementation of a $\nu$-OTDR  over $216.95\ \mathrm{km}$-long optical fiber. It is based on an ultra-low noise up-conversion single photon detector, and the NEP of the detector is suppressed to -139.7 $\mathrm{dBm/\sqrt{Hz}}$ by using long wavelength pump technology and a VBG as a narrow band filter. We also use laser pulses of $23\ \mathrm{dBm}$ peak power to reduce the measurement time. This apparatus can achieve a dynamic range of $42.19\ \mathrm{dB}$ and distance resolution of about $10\ \mathrm{cm}$ at the distance of $216.95\ \mathrm{km}$ in measurement time of $13\ \mathrm{minutes}$.

\section{Acknowledge}
The authors acknowledge Jun Zhang, Yang Liu, Yan-Ping Chen, Han Zhang, Tian-Ming Zhao and Xiu-Xiu Xia for their useful discussions. This work has been supported by the National Fundamental Research Program (under Grant No. 2011CB921300 and 2011CBA00300), the NNSF of China, the CAS, and the Shandong Institute of Quantum Science $\&$ Technology Co., Ltd. J.S.P. and M.M.F. acknowledge the U.S. AFOSR for their support under Grant No. FA9550-09-1-0233.

\end{document}